\documentclass{Interspeech2024}
\usepackage{color,amsmath,url,times,tabularx,bbm,amssymb,multirow}
\usepackage{footnote}
\makesavenoteenv{tabular}
\makesavenoteenv{table}
\usepackage{makecell} % 引入makecell包
\usepackage{cite}

\usepackage{graphicx}
\usepackage{bm}
\interspeechcameraready

\title{Audio-text Retrieval with Transformer-based Hierarchical Alignment and Disentangled Cross-modal Representation}

\name[affiliation={}]{Yifei}{Xin}
\name[affiliation={}]{Zhihong}{Zhu}
\name[affiliation={}]{Xuxin}{Cheng}
\name[affiliation={}]{Xusheng}{Yang}
\name[affiliation={*}]{Yuexian}{Zou}

\address{School of ECE, Peking University, China \thanks{This paper was partially supported by NSFC (No: 62176008) and Shenzhen Science \& Technology Research Program (No:GXWD20201231165807007-20200814115301001).}\thanks{$^{*}$ Yuexian Zou is the corresponding author.}}
\email{xinyifei@stu.pku.edu.cn}

\begin{document}

\maketitle
\begin{abstract}
Most existing audio-text retrieval (ATR) approaches typically rely on a single-level interaction to associate audio and text, limiting their ability to align different modalities and leading to suboptimal matches. In this work, we present a novel ATR framework that leverages two-stream Transformers in conjunction with a Hierarchical Alignment (THA) module to identify multi-level correspondences of different Transformer blocks between audio and text. Moreover, current ATR methods mainly focus on learning a global-level representation, missing out on intricate details to capture audio occurrences that correspond to textual semantics. To bridge this gap, we introduce a Disentangled Cross-modal Representation (DCR) approach that disentangles high-dimensional features into compact latent factors to grasp fine-grained audio-text semantic correlations. Additionally, we develop a confidence-aware (CA) module to estimate the confidence of each latent factor pair and adaptively aggregate cross-modal latent factors to achieve local semantic alignment. Experiments show that our THA effectively boosts ATR performance, with the DCR approach further contributing to consistent performance gains.

\end{abstract}

\section{Introduction}
The audio-text retrieval (ATR) task is to retrieve items from one modality when provided with a caption or audio clip from another. To align audio and text modalities, deep neural networks \cite{oncescu2021audio,xin23d_interspeech,zhao24h_interspeech,xin2022audio,cheng2024soul} have been leveraged to extract audio and text representations, which are subsequently mapped to a joint latent space to measure the relevance of audio-text pairs. However, many existing ATR methods adopt two-stream encoders with different architectures: for instance, CNNs for audio and Transformers for text \cite{xin24_interspeech,xin2023pooling}. Such architectural disparities may lead to varied semantic distribution spaces, potentially weakening the audio-text alignment. Moreover, most existing ATR strategies \cite{mei2022metric,wu2022text} predominantly emphasize single-level alignment across two independent streams, neglecting multi-level correspondences of different modalities. It's notable that semantics across different modalities are multifaceted. Incorporating information from another modality can lead to better semantic representations.

To fill the above research gaps, we present a novel ATR framework based on dual-stream Transformers, composed of an audio Transformer, a text Transformer, and a Transformer-based hierarchical alignment (THA) module. With such unified architecture, the dual-stream encoders could yield features with more similar characteristics for audio and text, facilitating more straightforward audio-text interactions and alignments. As prior studies have pointed out that shallow blocks in Transformers tend to capture more static and local semantics while deeper blocks encompass more contextual information \cite{de2020s,van2019does}, our THA module enables the ATR network to learn to grasp the rich correspondences between audio and text across different semantic levels.

Besides, current ATR methods mainly concentrate on establishing a global alignment across modalities \cite{mei2022language,deshmukh2022audio}, which treats discriminative information equally and curtails the potential to capture fine-grained audio-text correlations. A more refined strategy would involve aligning text and audio based on local semantic clues, emulating human cognition and reasoning process. In pursuit of extracting salient cross-modal information, we employ disentangled representation learning \cite{wang2021self,liu2022disentangled} to separately encode semantic information about each latent variable in only a few dimensions. 

Building upon the above understanding, we introduce a disentangled cross-modal representation (DCR) approach for ATR, which disentangles high-dimensional audio and text representations into compact latent factors that purposefully capture fine-grained audio-text semantic relationships. Our optimization of latent factors is twofold: inter-factor and intra-factor. For the former, we minimize mutual information among different latent factors to identify representation subspaces with minimal relevance to each other for disentangling representation. For the latter, we maximize mutual information for each latent factor pair separately, facilitating local alignment between corresponding audio and text pair. In order to further enhance the semantic matching of cross-modal information contained in audio-text pairs, we introduce a confidence-aware (CA) module to estimate the confidence of each cross-modal latent factor pair. Lastly, we utilize the confidence as the weight to adaptively aggregate all factor pairs to compute the text-audio similarity.

The primary contributions of this paper include: 
\begin{itemize} 
\item We present an ATR framework that leverages two-stream Transformers combined with a Hierarchical Alignment (THA) module which comprehensively explores multi-level correspondences of different Transformer blocks between audio and text.
\item We introduce a disentangled cross-modal representation (DCR) approach integrated with a confidence-aware (CA) module for ATR, which disentangles high-dimensional audio and text features into compact latent factors to capture fine-grained audio-text relationships and achieve local semantic alignment.
\item Experiments demonstrate that our THA significantly improves the ATR performance. Moreover, our DCR also consistently boosts the performance to a great extent.
\end{itemize}

\section{Proposed Methods}
\label{sec:proposed method}
\subsection{Dual-stream Transofrmers}
\noindent\textbf{Text Transformer.} Given a text sentence $T$ comprising $N$ words, we leverage BERT \cite{kenton2019bert} to extract word-level representations $T^l=\left\{t_n^l \mid t_1^l, t_2^l, \ldots, t_N^l\right\}$, where $l$ corresponds to the $l$-th Transformer block of BERT. In our approach, we utilize BERT-base with 12 Transformer blocks. 

\noindent\textbf{Audio Transformer.} Instead of the common practice of leveraging CNN-based architectures like ResNet38 \cite{mei2022metric,kong2020panns} for audio feature extraction, we utilize HTSAT \cite{chen2022hts} as the audio encoder, employing Swin Transformer \cite{liu2021swin} as the backbone, which demonstrates great potential and surpasses CNN models in various audio-related tasks \cite{chen2022hts}. Given an audio spectrogram $A \in \mathbb{R}^{H \times W \times 1}$, we first split the audio spectrogram into several non-overlapping patches, each considered a token to feed into the Transformer blocks. We take the output features of multiple stages as hierarchical audio representations to convey richer semantics, and represent the output frame-level features of each stage as $A^i$, where $i$ denotes the $i$-th stage. Due to the computational intensity of stage 1's large token numbers, we discard the output feature of stage 1. Outputs from stages 2-4 are labeled as low-, middle-, and high-level semantic audio representations. Thus, the overall audio hierarchical representation $A_h$ is denoted as: $A_h=\operatorname{HTSAT}(A)=\left\{A^{s2}, A^{s3}, A^{s4}\right\}$. To be consistent with BERT's block representation, we map stages in $A^{s2}, A^{s3}$, and $A^{s4}$ to their corresponding blocks, i.e., 4, 10, 12-th block in BERT, given that HTSAT includes 4 stages with 2, 2, 6, 2 swin-transformer blocks.

\subsection{Hierarchical Alignment Module}
%Given that both BERT and Swin Transformer comprise 12 layers, we symmetrically leverage the outputs of their same layers for multi-level semantic representations, i.e., outputs from the 4th, 10th, and 12th layers of both architectures are extracted for the hierarchical representations of audio and text.
After obtaining the frame-level and word-level representations, our Transformer-based hierarchical alignment (THA) integrates the three-level semantics to learn correlations between audio and text. As illustrated in Figure 1, we calculate the cosine similarities for each level and combine the hierarchical similarities by addition. Specifically, we implement text-enhanced and audio-enhanced stacked cross attention \cite{lee2018stacked,bin2023unifying,yin2023afl} respectively. Supposing the output of the $l$-th block of Swin Transformer $A^l=\left\{a_m^l \mid a_1^l, a_2^l, \ldots, a_M^l\right\}$ with $M$ audio patch tokens, and the output of the $l$-th block of BERT $T^l=\left\{t_t^l \mid t_1^l, t_2^l, \ldots, t_N^l\right\}$ with $N$ words, the text-enhanced stacked cross attention first calculates the similarities between all token-word pairs: $s_{mn}^l=\frac{a_m^{l^\mathsf{T}} t_n^l}{ {\lVert{a_m^l}\rVert}{\lVert{t_n^l}\rVert} }$, where $s_{mn}^l$ denotes the similarity between $m$-th token and $n$-th word in the $l$-th block of both HTSAT and BERT. Based on the token-word similarity, we aggregate the weighted words representations for each audio token to obtain the cross-modal fused text representations:
\begin{equation}
\hat{a}_m^l=\sum_{n=1}^N \alpha_{mn}^l t_n^l, \quad \alpha_{mn}^l=\frac{\exp \left(\lambda * \bar{s}_{mn}^l\right)}{\sum_{n=1}^N \exp \left(\lambda * \bar{s}_{mn}^l\right)}, 
\end{equation}
where $\lambda$ denotes the temperature hyper-parameter. $\bar{s}_{mn}^l$ is normalized similarity: $\bar{s}_{mn}^l=\frac{\left[s_{mn}^l\right]_{+}}{\sqrt{\sum_{m=1}^M\left[s_{mn}^l\right]_{+}^2}}$, where $[x]_{+}=\max (x, 0)$ denotes the hinge function to keep the similarity with positive value. Lastly, we compute the cosine similarity for each token-word pair $\left(a_m^l, \hat{a}_m^l\right)$, and aggregate all the pairs for the overall similarity:
$S_{THA}(A, T)=\sum_l \sum_m \frac{a_m^{l^T} \hat{a}_m^l}{\lVert a_m^l\rVert\lVert\hat{a}_m^l\rVert}$. Likewise, the audio-enhanced stacked cross attention follows similar patterns. 

After passing through all blocks of BERT and HTSAT, we also follow previous works \cite{mei2022metric,wu2022text} that take the [CLS] token output as the global text feature $\bar{t}$, and leverages an average pooling layer to aggregate frame-level embeddings into a global representation $\bar{a}$, which is followed by a dot-product operation between the two holistic representations to measure feature similarity: $S_{DP}(\bar{a}, \bar{t}) = \frac{\bar{a} \cdot \bar{t}}{{\Vert \bar{a} \Vert}_2 {\Vert \bar{t} \Vert}_2}$. Next, we will introduce a disentangled cross-modal representation (DCR) module to further improve the discriminative power of our ATR network for fine-grained cross-modal matching.
\begin{figure}[t]
  \centering
  \includegraphics[width=1.0\linewidth]{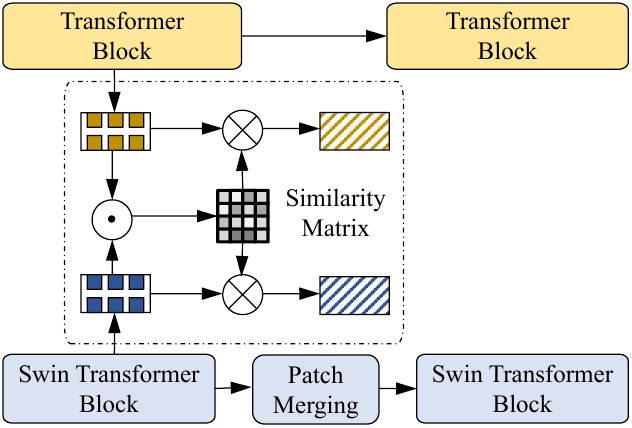}
  \caption{Illustration of our Transformer-based hierarchical alignment module. We take one-stage interaction as an example.}
  \vspace*{-\baselineskip}
\end{figure}

\subsection{Disentangled Cross-modal Representation from both Inter-factor and Intra-factor Perspectives}
Currently, most ATR methods \cite{mei2022language,deshmukh2022audio} yield holistic text representations and audio representations for similarity calculation. However, such global representations characterize semantic information of the input text and audio in an entangled state. As a result, direct similarity matching of these representations cannot capture fine-grained cross-modal correspondences between a set of diverse phrases in textual descriptions and a set of various audio events within audio clips. To address this, we learn discriminative disentangled latent factors from text and audio representations to explicitly measure local cross-modal correlations.

To initiate, we simplify by considering the sentence-level representation $\bar{t} \in \mathbb{R}^D$. According to the setting of disentangled representation learning \cite{yang2022cross,jin2023text}, we assume that the sentence-level representation can be decomposed into $K$ independent latent factors, i.e., $E^t=\left[e_1^t, e_2^t, \ldots, e_K^t\right]$. Each latent factor $e_k^t \in \mathbb{R}^{\frac{D}{K}}$ symbolizes a distinct textual semantic element. The independence of the latent factors ensures little semantic overlap among them. In this case, we project the text feature $\bar{t}$ into $K$ factors separately, and get the $k_{\text {th}}$ latent factor $e_k^t$: $e_k^t=W_k^t \bar{t}$, where $W_k^t \in \mathbb{R}^{\frac{D}{K} \times D}$ denotes a trainable parameter. Similarly, the latent factor $e_k^a$ of audio representation is calculated as: $e_k^a=W_k^a \bar{a}$. By combining the inter-factor decoupling and intra-factor alignment approaches introduced below, we explicitly map features to distinct representational subspaces corresponding to different semantic elements, empowering the model to optimize and reason cross-modal semantic information separately from these representation subspaces.
\begin{figure}[t]
  \centering
  \includegraphics[width=1.0\linewidth]{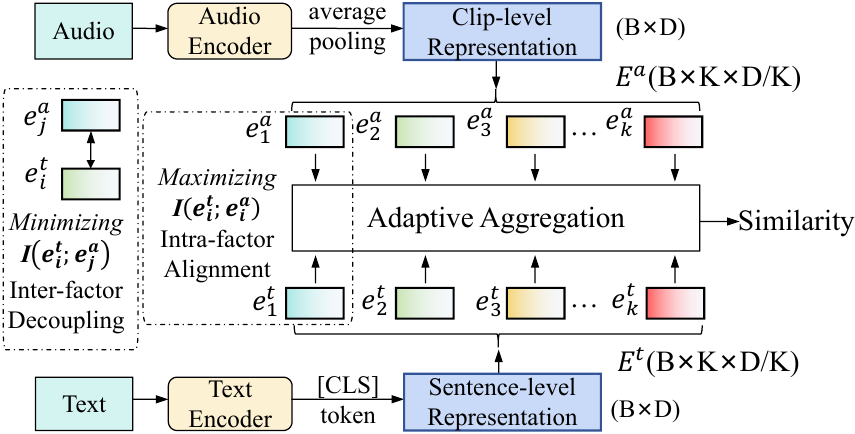}
  \caption{Overview of the disentangled cross-modal representation (DCR) framework.}
  \label{fig:adapt-pic}
  \vspace*{-\baselineskip}
\end{figure}

\noindent\textbf{Inter-factor Decoupling.} To identify representation subspaces with minimal relevance to each other and thus enhance the discriminative power for cross-modal semantic matching, we minimize the inter-factor mutual information. Taking two latent factors $e_i^t$ and $e_j^a$, their mutual information is captured via their probabilistic density functions as:
\begin{equation}
I\left(e_i^t ; e_j^a\right)=\mathbb{E}_{\boldsymbol{t}, \boldsymbol{a}}\left[p\left(e_i^t, e_j^a\right) \log \frac{p\left(e_i^t, e_j^a\right)}{p\left(e_i^t\right) p\left(e_j^a\right)}\right].
\end{equation}
Given the difficulty of directly quantifying mutual information, we follow previous work \cite{sreekar2021mutual,Do2020Theory,zhuang2024towards} to measure the mutual information implicitly. Specifically, for the latent factors $e^t \in \mathbb{R}^{\frac{D}{K}}$ and $e^a \in \mathbb{R}^{\frac{D}{K}}$, we first normalize them: $z^t=\frac{e^t-\mathbb{E}\left[e^t\right]}{\sqrt{\operatorname{Var}\left[e^t\right]}}, \quad z^a=\frac{e^a-\mathbb{E}\left[e^a\right]}{\sqrt{\operatorname{Var}\left[e^a\right]}}$,
where $z^t, z^a$ have the same mean and standard deviation. Then, we compute the covariance of $z^t$ and $z^a$: $C_{i, j}=\mathbb{E}_{\boldsymbol{t}, \boldsymbol{a}}\left[\left(z_i^t\right)^{\top} z_j^a\right]$, where $z_i^t$ and $z_j^a$ are the normalized features of $e_i^t$ and $e_j^a$, respectively. As the meaning of $p\left(e_i, e_j\right)$ is the probability that $e_i$ matches $e_j$, it can be expressed as:
\begin{equation}
\begin{aligned}
p\left(e_i, e_j\right) & =\frac{\mathbb{E}\left[\left(z_i\right)^{\top} z_j\right]}{\sum_{k=1}^K \mathbb{E}\left[\left(z_k\right)^{\top} z_j\right]}=\frac{C_{i, j}}{\sum_{k=1}^K C_{k, j}}.
\end{aligned}
\end{equation}
In this case, minimizing $p\left(e_i, e_j\right)$ is equivalent to the minimization of $C_{i, j}$. Therefore, our approach minimizes $I\left(e_i^t ; e_j^a\right)$ by minimizing the inter-factor decoupling loss $\mathcal{L}_D$ defined as: $\mathcal{L}_D=\sum_i \sum_{j \neq i}\left(C_{i, j}\right)^2$. This loss minimizes the mutual information for the negative pairs $\left(e_i^t, e_j^a\right)$, achieving the decoupling of latent factors.

\noindent\textbf{Intra-factor Alignment.} To effectively disentangle the $K$ latent factors from both text and audio representations, it's crucial to discern cross-modal semantics within text-audio correspondences. In this case, we optimize the representation subspace for each latent factor pair independently. Central to our strategy is treating each representation subspace separately to powerfully capture the latent factors' correlation within text-audio pairs. In detail, we maximize the mutual information between the matching text and audio latent factors within the identical subspace. The intra-factor alignment loss $\mathcal{L}_A$ is denoted as: $\mathcal{L}_A=\sum_i\left(1-C_{i, i}\right)^2$, which maximizes the mutual information of each positive pair $\left(e_i^t, e_i^a\right)$ independently.

\subsection{Adaptive Aggregation for Similarity Calculation}
To further enhance cross-modal semantic alignment for audio-text latent factors, we first present a confidence-aware (CA) module to estimate the confidence for each audio-text latent factor pair. Concretely, for the $i_{th}$ subspace, we concatenate the latent factor $e_i^t \in \mathbb{R}^{\frac{D}{K}}$ with the latent factor $e_i^a \in \mathbb{R}^{\frac{D}{K}}$ to form the input $\hat{e}_i=\left[e_i^t, e_i^a\right] \in \mathbb{R}^{\frac{2 D}{K}}$ for our CA module. The confidence score for the $i_{th}$ subspace is then derived using the function: $g_i=\operatorname{MLP}\left(\hat{e}_i\right)$, where MLP consists of two linear layers and a ReLU activation function. A smaller value of $g_i$ implies that the semantic element is less likely to match the $i_{th}$ subspace. Leveraging these confidence scores as weights, we adaptively aggregate all factor pairs to compute the text-audio similarity, defined as:
\begin{equation}
S_{DCR}=\sum_{i=1}^K g_i \frac{\left(e_i^t\right)^{\top} e_i^a}{\lVert e_i^t\rVert\lVert e_i^a\rVert}.
\end{equation}
Following common practice, we adopt a contrastive learning framework \cite{mei2022metric}, specifically utilizing the NT-Xent loss \cite{chen2020simple}, to optimize the cross-modal similarity:
\begin{equation}
\begin{aligned}
\mathcal{L}_S=-\frac{1}{B}\left(\sum_{i=1}^B \limits \log \frac{\exp \left(S_{i, i} / \tau\right)}{\sum_{j=1}^B \exp \left(S_{i, j} / \tau\right)}+\right. \\
\left. \sum_{i=1}^B \limits \log \frac{\exp \left(S_{i, i} / \tau\right)}{\sum_{j=1}^B \exp \left(S_{j, i} / \tau\right)}\right),
\end{aligned}
\end{equation}
where $B$ is the batch size, $\tau$ serves as a temperature hyper-parameter, $i$ and $j$ indicate the sample index within a batch. Notably, the overall similarity $S$ can be only $S_{DCR}$, or it can be a combination of multiple similarities (e.g., $S_{THA}+S_{DCR}$). Combining the aforementioned objectives for text-audio similarity $\mathcal{L}_S$, inter-factor decoupling $\mathcal{L}_D$ and intra-factor alignment $\mathcal{L}_A$, the whole training loss is $\mathcal{L}=\mathcal{L}_S+\alpha \mathcal{L}_D+\beta \mathcal{L}_A$, where $\alpha$ and $\beta$ are optimization hyper-parameters guiding the balance between individual loss components.
\begin{table*}
  \caption{ATR performance on the AudioCaps dataset.}
  \vspace*{-0.3cm}
  \centering
  \begin{tabular}{c|ccc|ccc}
    \toprule
    \multirow{2}{*}{Methods} & \multicolumn{3}{c|}{Text-to-Audio} & \multicolumn{3}{c}{Audio-to-Text} \\
    & \textbf{R@1} & \textbf{R@5} & \textbf{R@10} & \textbf{R@1} & \textbf{R@5} & \textbf{R@10}\\
    \midrule
    DP (ResNet38) \cite{mei2022metric} & 33.9$\pm$0.4 & 69.7$\pm$0.2 & 82.6$\pm$0.3 & 39.4$\pm$1.0 & 72.0$\pm$1.0 & 83.9$\pm$0.6\\
    \midrule    
    \textbf{DP (HTSAT)} & \textbf{38.3$\pm$0.2} & \textbf{72.2$\pm$0.1} & \textbf{83.2$\pm$0.5} & \textbf{45.1$\pm$0.6} & \textbf{77.7$\pm$0.8} & \textbf{88.1$\pm$0.6} \\
    \textbf{THA} & \textbf{39.9$\pm$0.3} & \textbf{73.6$\pm$0.7} & \textbf{84.9$\pm$0.1} & \textbf{46.8$\pm$0.4} & \textbf{79.1$\pm$0.2} & \textbf{89.4$\pm$0.5} \\   
    \textbf{THA+DP} & \textbf{40.1$\pm$0.4} & \textbf{73.9$\pm$0.2} & \textbf{85.2$\pm$0.3} & \textbf{47.1$\pm$0.1} & \textbf{79.4$\pm$0.4} & \textbf{89.6$\pm$0.4} \\   
    \textbf{DCR} & \textbf{40.6$\pm$0.3} & \textbf{74.1$\pm$0.8} & \textbf{84.3$\pm$0.5} & \textbf{46.6$\pm$0.2} & \textbf{79.3$\pm$1.0} & \textbf{89.9$\pm$0.9} \\
    \textbf{THA+DCR} & \textbf{41.1$\pm$0.6} & \textbf{74.9$\pm$0.8} & \textbf{85.7$\pm$0.1} & \textbf{47.8$\pm$0.1} & \textbf{80.6$\pm$0.4} & \textbf{90.6$\pm$0.3} \\
  \bottomrule
\end{tabular}
\vspace*{-\baselineskip}
\end{table*}
\section{Experiments}
\label{sec:exp}
\subsection{Dataset and Experiment Details}
We evaluate our methods on the AudioCaps \cite{kim2019audiocaps} and Clotho \cite{drossos2020clotho} datasets. AudioCaps contains around 50,000 audio samples; its training set consists of 49,274 samples each with one caption, while the validation and evaluation sets contain 494 and 957 samples, each with five captions. Clotho v2 includes 3,839 training, 1,045 validation, and 1,045 testing samples, each associated with five descriptive sentences. In this study, we follow the same pipeline in \cite{mei2022metric} to train our network. We perform experiments by fine-tuning the audio and text encoders. We set $K$ = 8, $\alpha$ = 0.01, $\beta$ = 0.005. We repeat experiments three times with different training seeds, presenting the mean and standard deviation of the metrics on the AudioCaps dataset. Recall at rank k (R@k) is utilized as the evaluation metric. R@1, R@5, and R@10 are presented in our results.
%, which quantifies the the fraction of targets ranking within the top-k outputs
\begin{table}\footnotesize
  \caption{ATR performance on the Clotho dataset.}
  \vspace*{-\baselineskip}
  \centering
  \label{tab:freq}
  \begin{tabular}{c|ccc|ccc}
    \toprule
    \multirow{2}{*}{Methods} & \multicolumn{3}{c|}{Text-to-Audio} & \multicolumn{3}{c}{Audio-to-Text} \\
    & \textbf{R@1} & \textbf{R@5} & \textbf{R@10} & \textbf{R@1} & \textbf{R@5} & \textbf{R@10}\\
    \midrule
    DP(HTSAT) & 11.4 & 31.9 & 43.1 & 14.0 & 32.4 & 45.4 \\
    \midrule
    \textbf{THA} & \textbf{13.2} & \textbf{33.4} & \textbf{44.9} & \textbf{15.1} & \textbf{33.7} & \textbf{46.6} \\   
    \textbf{THA+DP} & \textbf{13.7} & \textbf{33.9} & \textbf{45.2} & \textbf{15.4} & \textbf{33.8} & \textbf{46.7} \\   
    \textbf{DCR} & \textbf{14.1} & \textbf{34.2} & \textbf{45.5} & \textbf{15.9} & \textbf{34.2} & \textbf{46.9} \\
    \textbf{THA+DCR} & \textbf{14.8} & \textbf{34.9} & \textbf{46.7} & \textbf{16.8} & \textbf{35.6} & \textbf{47.8} \\
  \bottomrule
\end{tabular}
\vspace*{-\baselineskip}
\vspace*{-0.1cm}
\end{table}

\subsection{Experimental Results}
As shown in Table 1 and Table 2, we compare our proposed methods with previous baselines on the AudioCaps and Clotho datasets. “DP (ResNet38)” denotes the existing mainstream ATR approach leveraging ResNet38 \cite{kong2020panns} as the audio encoder and employing the NT-Xent loss \cite{chen2020simple} that performs the dot product operation to compute the global similarity between audio-text pairs, while “DP (HTSAT)” indicates that we adopt HTSAT \cite{chen2022hts} as the audio encoder, based on the Swin Transformer backbone. “THA” represents that we only use our Transformer-based hierarchical alignment method, while “THA+DP” refers to an integration of our THA module with the prevalent dot product strategy to enhance feature similarity computation. “DCR” denotes that we only leverage our disentangled cross-modal representation method, while “THA+DCR” combines our THA and DCR approaches.

It is clear that our methods consistently achieve better performance over the baselines both text-to-audio and audio-to-text retrieval tasks. Particularly, our THA+DCR combination stands out, delivering the highest scores across all evaluation metrics. This underlines the efficacy of combining the hierarchical alignment with our representation disentanglement approaches. When applied individually, both THA and DCR manifest commendable performance, which not only demonstrates the potency of each method but also underscores their potential as standalone techniques. Moreover, the marginal improvements observed in THA+DP over THA suggest that integrating existing techniques like dot product operations can further refine our methods, indicating the compatibility of our approaches.

\begin{table}\footnotesize
  \caption{Effect of each part of our DCR method.}
  \vspace*{-\baselineskip}
  \centering
  \begin{tabular}{c|ccc|ccc}
    \toprule
    \multirow{2}{*}{Method} & \multicolumn{3}{c|}{Text-to-Audio} & \multicolumn{3}{c}{Audio-to-Text} \\
    & \textbf{R@1} & \textbf{R@5} & \textbf{R@10} & \textbf{R@1} & \textbf{R@5}  & \textbf{R@10}\\
    \midrule      
    DP(HTSAT) & 38.3 & 72.2 & 83.2 & 45.1 & 77.7 & 88.1\\   
    + $\mathcal{L}_D$ & \textbf{39.1} & \textbf{72.8} & \textbf{83.6} & \textbf{45.9} & \textbf{78.2} & \textbf{88.7}\\ 
    + $\mathcal{L}_A$ & \textbf{39.4} & \textbf{73.3} & \textbf{83.9} & \textbf{46.2} & \textbf{78.6} & \textbf{89.1}\\
    + CA & \textbf{40.6} & \textbf{74.1} & \textbf{84.3} & \textbf{46.6} & \textbf{79.3} & \textbf{89.9}\\
  \bottomrule
\end{tabular}
\vspace*{-\baselineskip}
\vspace*{-0.3cm}
\end{table}

\subsection{Ablation Study}
In this part, we discuss the influences of each part of our DCR method, the number of latent factors, and the selection of hyper-parameters $\alpha$ and $\beta$ on the AudioCaps dataset.

\textbf{Each part of our DCR method.} Table 3 studies the importance of each part of our DCR method. We find that compared with the intra-factor alignment ($\mathcal{L}_A$) of latent factors, the improvements of the inter-factor decoupling ($\mathcal{L}_D$) is more notable. This is likely ascribed to the efficacy of the inter-factor decoupling strategy in mitigating the interference of mismatched semantic elements. Additionally, the integration of our confidence-aware (CA) module yields marked improvement, attributed to its adaptive mechanism in determining the text-audio similarity for local cross-modal matching. Finally, our full method obtains the best performance, thus strongly showing that the three components are all beneficial for aligning audio and text representations.
\begin{table}\footnotesize
  \caption{Effect of the number of latent factors.}
  \vspace*{-\baselineskip}
  \centering
  \begin{tabular}{c|ccc|ccc}
    \toprule
    \multirow{2}{*}{Method} & \multicolumn{3}{c|}{Text-to-Audio} & \multicolumn{3}{c}{Audio-to-Text} \\
    & \textbf{R@1} & \textbf{R@5} & \textbf{R@10} & \textbf{R@1} & \textbf{R@5}  & \textbf{R@10}\\
    \midrule      
    6 & 40.3 & 73.7 & 84.0 & 46.4 & 78.9& 89.5\\   
    8 & \textbf{40.6} & \textbf{74.1} & \textbf{84.3} & \textbf{46.6} & \textbf{79.3} & \textbf{89.9}\\ 
    10 & 40.2 & 73.8 & 83.8 & 46.3 & 78.9 & 89.8\\
  \bottomrule
\end{tabular}
\vspace*{-\baselineskip}
\vspace*{-0.1cm}
\end{table}

\begin{table}\footnotesize
  \caption{Effect of the trade-off hyper-parameters $\alpha$ and $\beta$.}
  \vspace*{-\baselineskip}
  \centering
  \begin{tabular}{c|ccc|ccc}
    \toprule
    \multirow{2}{*}{Method} & \multicolumn{3}{c|}{Text-to-Audio} & \multicolumn{3}{c}{Audio-to-Text} \\
    & \textbf{R@1} & \textbf{R@5} & \textbf{R@10} & \textbf{R@1} & \textbf{R@5}  & \textbf{R@10}\\
    \midrule      
    $\alpha$ = 0.005 & 40.3 & 74.0 & 84.2 & 46.6 & 78.9 & 89.5\\   
    $\alpha$ = 0.01 & \textbf{40.6} & \textbf{74.1} & \textbf{84.3} & \textbf{46.6} & \textbf{79.3} & \textbf{89.9}\\ 
    $\alpha$ = 0.02 & 40.4 & 73.7 & 84.2 & 46.3 & 79.1 & 89.7\\
    \midrule   
    $\beta$ = 0.001 & 40.2 & 73.9 & 83.9 & 46.2 & 79.3 & 89.6\\   
    $\beta$ = 0.005 & \textbf{40.6} & \textbf{74.1} & \textbf{84.3} & \textbf{46.6} & \textbf{79.3} & \textbf{89.9}\\  
    $\beta$ = 0.01 & 40.5 & 74.1 & 84.1 & 46.4 & 79.2 & 89.8\\
  \bottomrule
\end{tabular}
\vspace*{-\baselineskip}
\vspace*{-0.3cm}
\end{table}

\textbf{The number of latent factors.} In Table 4, we show the results of different numbers of latent factors $E=\left[e_1, e_2, \ldots, e_K\right]$, where $e_k \in \mathbb{R}^{\frac{D}{K}}$. We find that $K$ = 8 achieves the optimal performance. A smaller set of latent factors potentially constrains the network's ability to exploit fine-grained information. Conversely, an excessive number of latent factors diminishes the dimension of each latent factor $e_k$, thus compromising the discriminative ability of the factors.

\textbf{The selection of hyper-parameters $\alpha$ and $\beta$.} The hyper-parameters $\alpha$ and $\beta$ indicate the importance of $L_D$ and $L_A$ respectively. As shown in Table 5, we find that the optimal performance is achieved when $\alpha$ is set to 0.01. This suggests a slight but notable improvement over the settings where $\alpha$ equals 0.005 and 0.02. A similar pattern is observed for the $\beta$ parameter, where setting $\beta$ to 0.005 yields the best results.

\section{Conclusions}
\label{sec:conclusion}
In this work, we present an ATR framework that leverages two-stream Transformers combined with a hierarchical alignment (THA) module to unify the architectures of audio and text encoders, and enhance multi-level correspondences between audio and text. Furthermore, we introduce a disentangled cross-modal representation (DCR) approach that disentangles high-dimensional features into compact latent factors and adaptively aggregates cross-modal latent factors to achieve fine-grained local semantic alignment. Experiments show that our THA effectively boosts the ATR performance, with the DCR approach further enhancing these improvements.
\bibliographystyle{IEEEtran}
\bibliography{mybib}

\end{document}